\documentclass[12pt]{article}

\usepackage{scicite}
\usepackage{times}

\usepackage{color,gensymb,amsmath,amssymb,graphicx,morefloats}
\usepackage{graphicx,morefloats}
\usepackage[]{multibib}
\usepackage{subfigure} 
\usepackage{color,soul}
\usepackage{caption}
\newcites{SM}{Supplementary References}

\begin{document}

\thispagestyle{empty}

\hyphenation{va-ni-sh-ing de-lo-ca-li-za-tion}



\baselineskip24pt

\begin{center}

{\Large Singular charge fluctuations at a magnetic quantum critical
point}\\[0.4cm]

\normalsize{L.\ Prochaska$^{1,\ast}$, X.\ Li$^{2,\ast}$, D.\ C.\
MacFarland$^{1,3,\ast}$, A.\ M.\ Andrews$^3$,\\ M.\ Bonta$^4$, E.\ F.\
Bianco$^5$, S.\ Yazdi$^{6,\$}$, W.\ Schrenk$^7$, H.\ Detz$^7$, A.\
Limbeck$^4$,\\ Q.\ Si$^8$, E.\ Ringe$^6$,  G.~Strasser$^{3,7}$, J.\
Kono$^{2,6,8}$, and S.\ Paschen$^{1,8,\ddagger}$}\\[0.4cm]

\small\it{$^1$Institute of Solid State Physics, TU Wien, Wiedner Hauptstr.\ 8-10, 1040 Vienna, Austria\\[0.1cm]

$^2$Department of Electrical and Computer Engineering, 6100 Main Street, Rice University, Houston, Texas 77005, USA\\[0.1cm]

$^3$Institute of Solid State Electronics, TU Wien, Nanocenter Campus Gu{\ss}haus, Gu{\ss}hausstr.\ 25-25a, Geb\"aude CH, 1040 Vienna, Austria\\[0.1cm]

$^4$Institute of Chemical Technologies and Analytics, TU Wien, Getreidemarkt 9, 1060 Vienna, Austria\\[0.1cm]

$^5$Department of Chemistry, 6100 Main Street, Rice University, Houston, Texas 77005, USA\\[0.1cm]

$^6$Department of Materials Science and Nanoengineering, 6100 Main Street, Rice University, Houston, Texas 77005, USA\\[0.1cm]

$^7$Center for Micro- and Nanostructures, Nanocenter Campus Gu{\ss}haus, Gu{\ss}hausstr.\ 25-25a, Geb\"aude CH, 1040 Vienna, Austria\\[0.1cm]

$^8$Department of Physics and Astronomy, Center for Quantum Materials, 6100 Main Street, Rice University, Houston, Texas 77005, USA}

\vspace{0.4cm}

\today

\end{center}
\vspace{-0.2cm}

{\bf Strange metal behavior is ubiquitous in correlated materials ranging from
cuprate superconductors to bilayer graphene. There is increasing recognition
that it arises from physics beyond the quantum fluctuations of a Landau order
parameter which, in quantum critical heavy fermion antiferromagnets, may be
realized as critical Kondo entanglement of spin and charge. The dynamics of the
associated electronic delocalization transition could be ideally probed by
optical conductivity, but experiments in the corresponding frequency and
temperature ranges have remained elusive. We present terahertz time-domain
transmission spectroscopy on molecular beam epitaxy-grown thin films of
YbRh$_2$Si$_2$, a model strange metal compound. We observe frequency over
temperature scaling of the optical conductivity as a hallmark of beyond-Landau
quantum criticality. Our discovery implicates critical charge fluctuations as
playing a central role in the strange metal behavior, thereby elucidating one of
the longstanding mysteries of correlated quantum matter.}

\newpage




\noindent$^{\ast}$These authors contributed equally to this work.

\noindent$^{\$}$Present address:
Renewable and Sustainable Energy Institute, University of Colorado Boulder, Boulder, Colorado 80309, USA

\noindent$^{\ddagger}$To whom correspondence should be addressed; E-mail:  paschen@ifp.tuwien.ac.at.

\newpage

Quantum critical behavior as prescribed by the Landau framework of order
parameter fluctuations \cite{Her76.1,Mil93.1} has been clearly identified in
insulating quantum magnets such as LiHoF$_4$ \cite{Bit96.1} and TlCuCl$_3$
\cite{Mer14.1}. In strongly correlated metals, however, this framework often
fails. This hints at the charge carriers as giving rise to new forms of quantum
criticality. In the strange metal \cite{Gir18.1} regime of various correlated
systems \cite{Bru13.1}, electronic localization-delocalization transitions have
been evidenced
\cite{Pas04.1,Fri10.2,Cus12.1,Luo14.1,Bal03.1,Oik15.1,Bad16.1,Cao18.1} and it is
an outstanding question whether they are a key ingredient of beyond-Landau
quantum criticality. To make progress, it is essential to study the dynamics of
charge carriers in a suitable setting.

We chose the heavy fermion metal YbRh$_2$Si$_2$ \cite{Tro00.2} for our
investigation because it has a well-defined quantum critical point
\cite{Tro00.2,Geg02.1} and shows evidence for an electron
delocalization transition \cite{Pas04.1,Fri10.2} in its strange
metal regime. An ideal tool for such studies are optical conductivity
measurements in the relevant frequency window, which is typically the terahertz
(THz) range and below for heavy fermion systems. However, such measurements are
challenging on bulk samples because the Kramers-Kronig transformation to extract
the real and imaginary parts of the optical conductivity from reflectivity
measurements introduces substantial uncertainty at low frequencies
\cite{Ore90.1}. Thus, we resorted to an entirely different and previously
unattempted approach: We performed THz time-domain transmission spectroscopy
experiments on thin films of YbRh$_2$Si$_2$ grown by molecular beam epitaxy
(MBE). Our measurements reveal $\omega/T$ scaling of the optical conductivity,
thereby elucidating the mechanism for strange metal phenomena.

To grow epitaxial thin films of YbRh$_2$Si$_2$ on (THz transparent) Ge
substrates (Fig.\,\ref{MBEfilm}A), we used a specially equipped MBE system
\cite{YRS18SM}. The epitaxial growth of phase-pure YbRh$_2$Si$_2$ is confirmed
by X-ray diffraction (Fig.\,\ref{MBEfilm}B) \cite{YRS18SM}, and the high quality
of the film and the film-substrate interface are revealed by high-resolution
transmission electron microscopy (Fig.\,\ref{MBEfilm}C,\,D) \cite{YRS18SM}. The
temperature dependence of the (quasi) \cite{YRS18SM} dc electrical resistivity
$\rho(T)$ of these films \cite{YRS18SM} is similar to that of bulk single
crystals \cite{Tro00.2,Wes09.1} (Fig.\,\ref{rhoT}). Most notably, it displays
strange metal behavior, $\rho = \rho_0 + A' T^{\alpha}$ (Fig.\,\ref{rhoT}B),
with an exponent $\alpha$ that strongly deviates from the Fermi liquid value
$\alpha = 2$, and tends to $\alpha = 1$ in the low-temperature limit
(Fig.\,\ref{rhoT}C).

\begin{figure}[ht!]
\vspace{-1cm}

{\large\bf\hspace{-0.2cm} A \hspace{7.8cm} B}
\vspace{-0.2cm}

\centering
\subfigure{\includegraphics*[width=0.53\textwidth]{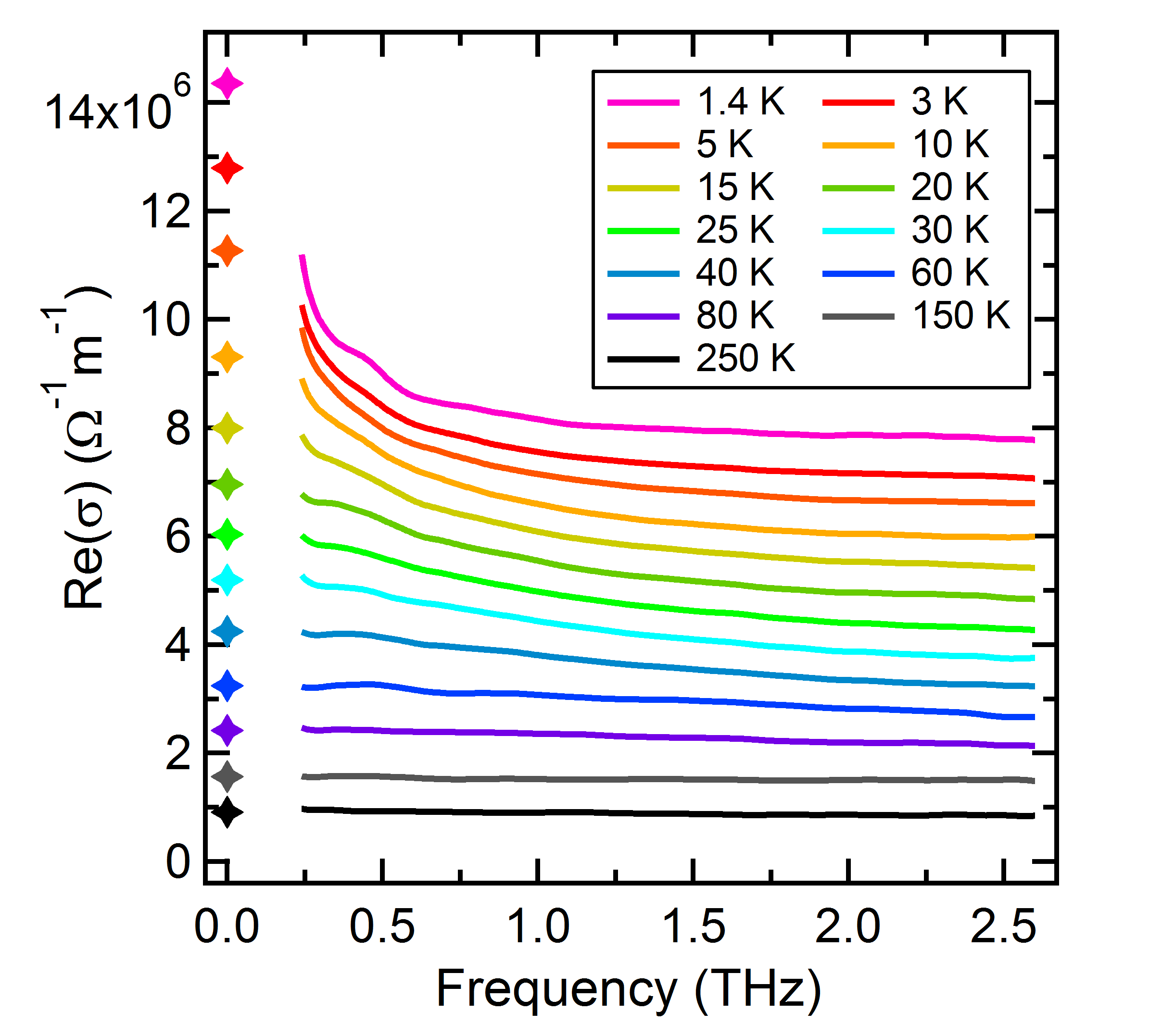}\hspace{-0.3cm}\includegraphics*[width=0.53\textwidth]{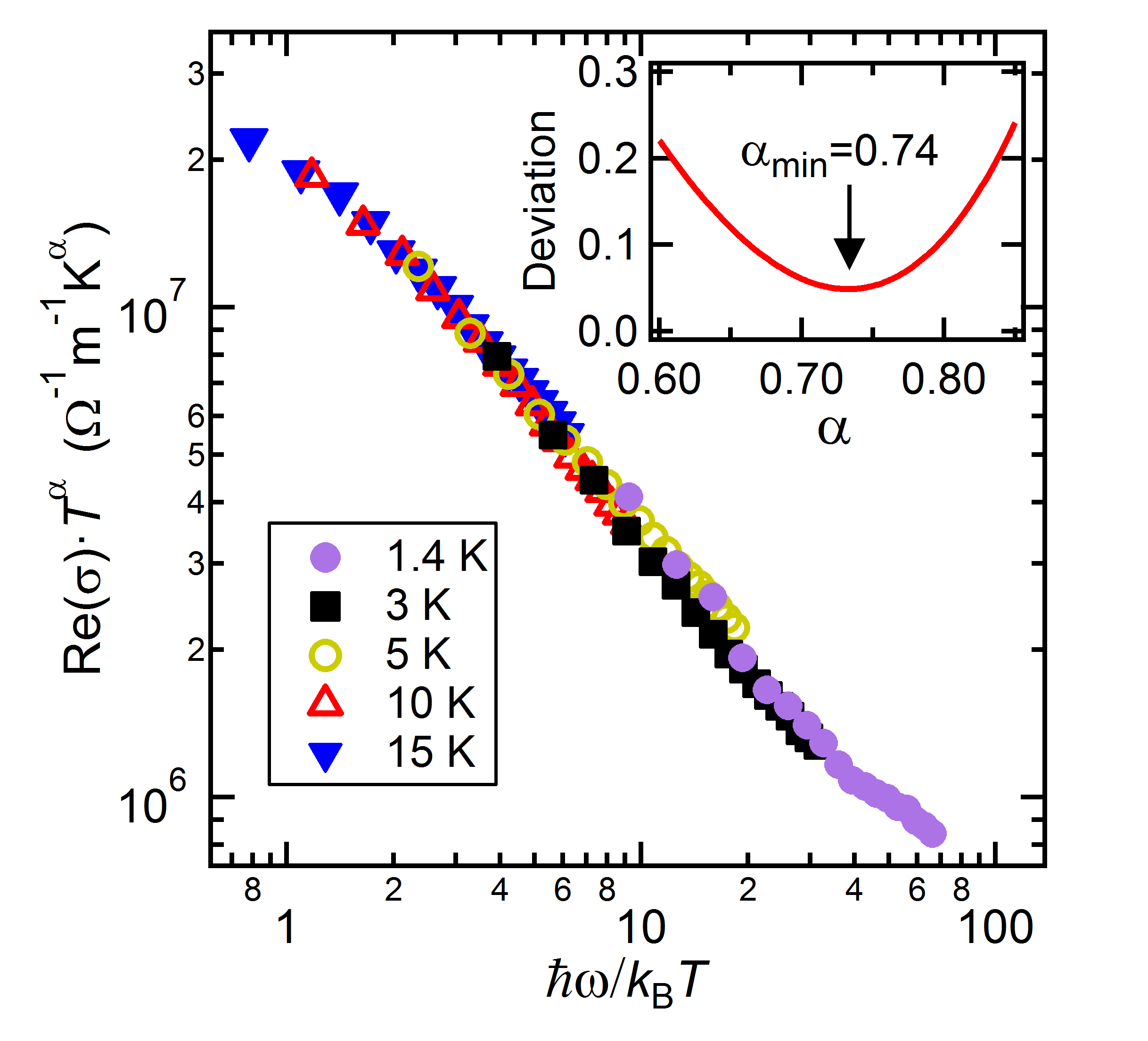}}
\vspace{0.5cm}


\caption{\label{sigmaTHz}
\baselineskip24pt
{\bf THz time-domain transmission spectroscopy of MBE-grown YbRh$_2$Si$_2$.}
({\bf A}) Real part of optical conductivity Re$(\sigma)$ vs frequency at
different temperatures, with corresponding dc values marked as zero-frequency
points. Curves below 250\,K are successively offset by $6 \times
10^5$\,$\Omega^{-1}m^{-1}$ for clarity. ({\bf B}) $\omega/T$ scaling, with a
critical exponent of $\alpha=0.74$, revealed by curves of
Re$[\sigma(\omega)]\cdot T^\alpha$ as a function of $\hbar\omega/(k_{\rm{B}}T)$
collapsing onto a single curve for temperatures $T\leq15$\,K and frequencies
below 2\,THz. ({\bf B} inset) Normalized deviation between curves as a function
of $\alpha$, revealing best scaling for $\alpha=0.74$.}
\end{figure}

The frequency dependence of the real part of the complex optical conductivity,
Re$(\sigma)$, measured at temperatures between 1.4 and 250\,K and frequencies
between 0.25 and 2.6\,THz, is shown in Fig.\,\ref{sigmaTHz}A (see Fig.\,S1 for
the imaginary part, Im$(\sigma)$). The dc electrical conductivity
$\sigma=1/\rho$ values, plotted as symbols at $\omega=0$, are compatible with
the extrapolation of the finite frequency results to zero frequency. Both
Re$(\sigma)$ and Im$(\sigma)$ are flat and featureless at temperatures above
about 80\,K, indicating strong incoherent scattering of charges, but develop
sizeable temperature and frequency dependence at lower temperatures, with
spectral weight of Re$(\sigma)$ being transferred to low frequencies. The
increasingly sharp and pronounced resonance of Re$(\sigma)$, with non-Lorentzian
shape, may, in clean samples, be associated with non-Fermi liquid behavior. In
the far-infrared range, where there is overlap with data from optical
reflectivity measurements on bulk single crystals \cite{Kim06.2}, the agreement
is good and our results confirm the deviations from a simple Drude response
discussed there.

To explore dynamical scaling, we plot Re$[\sigma(\omega)]\cdot T^\alpha$ as a
function of $\hbar\omega/(k_{\rm{B}}T)$ for temperatures ($T\le 15$\,K) well
below the material's Kondo temperature $T_{\rm{K}} = 24$\,K \cite{Tro00.2}
(Fig.\,\ref{sigmaTHz}B) and frequencies below 2\,THz. Indeed, for $\alpha=0.74$,
all curves collapse, demonstrating $\omega/T$ scaling of Re$[\sigma(\omega)]$.
This is the first time $\omega/T$ scaling of the charge carrier response is
directly observed in YbRh$_2$Si$_2$ and, to the best of our knowledge, in any
optical conductivity measurement on quantum critical heavy fermion compounds.

The critical exponent $\alpha$ is somewhat smaller than that of the dc
electrical resistivity in the $T\rightarrow 0$ limit, which is close to 1
(Fig.\,\ref{rhoT}C). Note, however, that both our thin film and bulk single
crystals \cite{Tro00.2,Wes09.1} show deviations from 1 at somewhat higher
temperatures. In particular, the exponent falls below 0.9 above about 8\,K for
our thin film (Fig.\,\ref{rhoT}C). Indeed, when limiting the scaling to
temperatures below 5\,K, we obtain $\alpha=0.83$ (Fig.\,S2). If, in addition, we
limit the frequency range to below 0.7\,THz, which corresponds to a thermal
energy of about 5\,K \cite{YRS18SM}, the exponent reaches 0.94. Given that
$1/\rho$ is the $\omega\rightarrow 0$ limit of Re$(\sigma)$, the exponent in the
dynamical scaling function of Re$(\sigma)$ should be compatible with that of
$\rho(T)$. Within the error bars of about 10\% for the experimental
determination of $\alpha$, this is indeed the case.

\begin{figure}[ht!]
\vspace{-1cm}

\centering
\subfigure{\includegraphics*[width=0.7\textwidth]{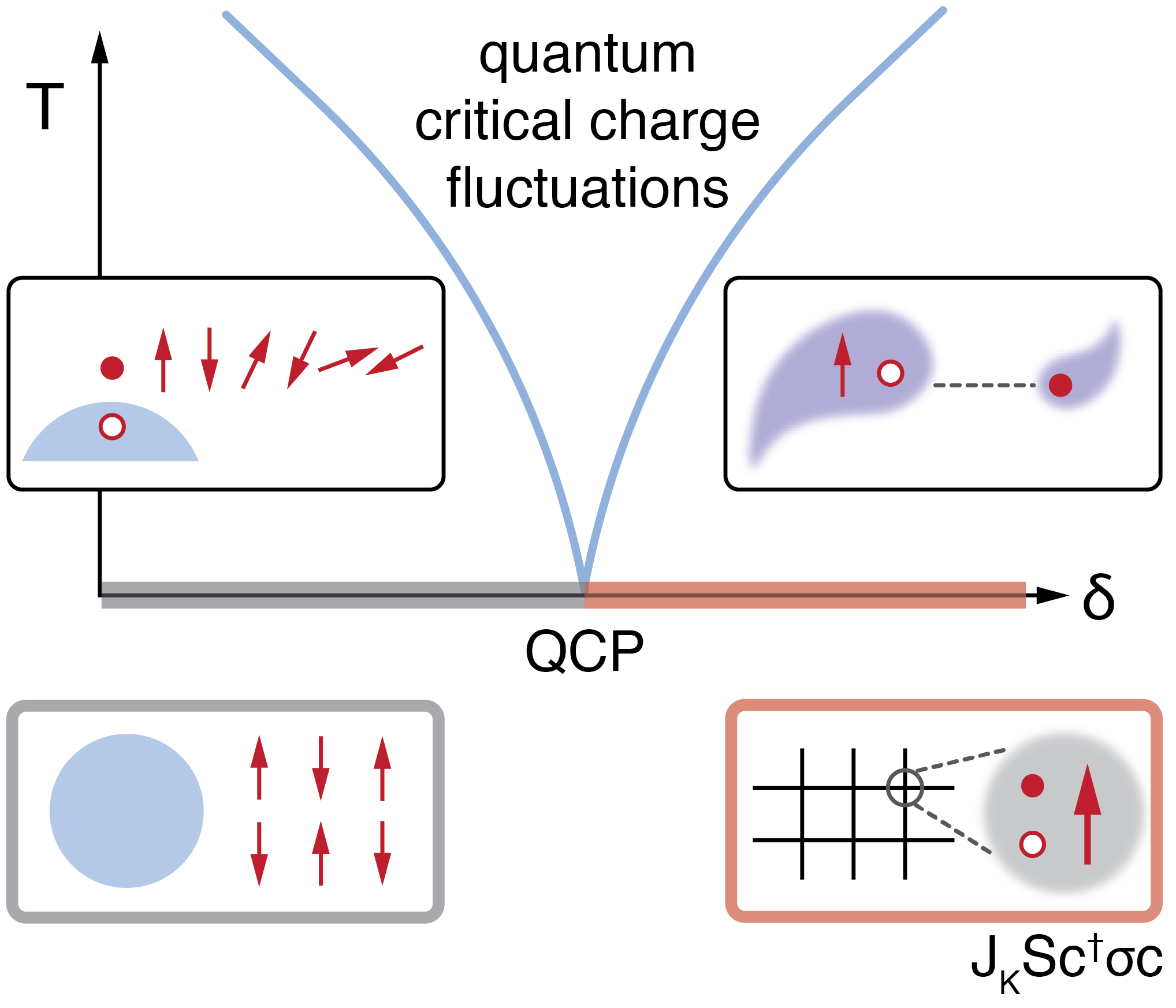}}
\vspace{0.5cm}


\caption{\label{sketch}
\baselineskip24pt
{\bf Sketch of quantum critical charge fluctuations emerging from Kondo
disentanglement.} Tuning a heavy fermion metal with a non-thermal parameter
$\delta$, which microscopically corresponds to the ratio of Kondo to RKKY
coupling, from an antiferromagnetic ground state with local moment order (bottom
left box; circle and arrows represent Fermi sphere and local moments,
respectively) to a Kondo entangled paramagnet (bottom right box; the
antiferromagnetic Kondo exchange $J_K$ favors the formation of a Kondo singlet
between the local moment $S$, represented as arrow, and the spin of the
conduction electrons $c^{\dagger} \sigma c$, i.e., the particle-hole excitation
of the Fermi sea in the spin-triplet channel) creates quantum critical charge
fluctuations in the excitation spectrum (top boxes, see main text).}
\end{figure}

How can the optical conductivity, which probes charge fluctuations, show
$\omega/T$ scaling at an antiferromagnetic quantum critical point where, {\it a
priori}, only spin fluctuations are expected -- and indeed observed
\cite{Aro95.1,Sch00.1,Sto12.1} -- to be critical? A natural way for this to
happen is to have a critical form of the Kondo entanglement between the local
moments and the conduction electrons \cite{Si01.1,Col01.1,Sen04.1}, as
illustrated in Fig.\,\ref{sketch}. Across the quantum critical point, the
conduction electrons go from being (asymptotically) decoupled from the local
moments (bottom left box) to being entangled with them (bottom right box).
Correspondingly, the elementary excitations change from single conduction
electrons/holes and separated spin excitations (top left box) to the heavy
quasiparticles (top right box), hybrids of the slow composite fermions (large
tadpole) and the bare conduction electrons (small tadpole). In other words, the
single electron excitations capture the continuous onset of the Kondo
entanglement at the quantum critical point and are part of the critical degrees
of freedom. Thus, optical conductivity, which probes the charge current of the
elementary excitations, manifests the singular fluctuations of the QCP. In fact,
within the Landau description of a metallic antiferromagnetic quantum critical
point \cite{Her76.1,Mil93.1}, the slow long-wavelength fluctuations of the order
parameter alone describe a Gaussian fixed point, where $\omega/T$ scaling is
violated. The incorporation of the single electron excitations in the quantum
critical spectrum not only makes charge fluctuations part of the quantum
criticality, but also turns the fixed point into an interacting one
\cite{Si01.1}, thereby leading to $\omega/T$ scaling.

Dynamical scaling of the optical conductivity in the region of $T$-linear
resistivity has also been analysed in an optimally doped Bi-2212 cuprate
\cite{Mar03.1}. There, different scaling functions are needed in different
$\omega/T$ ranges, leaving it open how the fluctuations of the charge carriers
connect with the robust linear-in-temperature resistivity of the cuprate
superconductors. By contrast, in the present study of YbRh$_2$Si$_2$, a single
$\omega/T$ scaling form is uncovered in its strange metal regime. It is
important to explore the dynamical scaling of the optical conductivity in other
materials classes with strange metal behavior; one can then assess whether the
charge carrier dynamics emerging from a localization-delocalization quantum
critical point, as proposed here, is a universal mechanism of strange metal
behavior. Our scaling results also provide an intriguing link to the quantum
scaling of metal-insulator transitions, both in Mott-Hubbard
\cite{Sen08.1,Ter11.1,Fur15.1} and in disordered systems \cite{Lee00.1}.

To summarize, we have discovered the dynamical scaling of charge fluctuations at
an antiferromagnetic quantum critical point. This long-sought-after effect was
made possible by mastering the challenging MBE thin-film
growth of a strange metal, which allowed for the application of THz time-domain
transmission spectroscopy into the quantum critical frequency range. Our results
demonstrate the charge carriers to be a central ingredient of the singular
physics at the border of antiferromagnetic order, thereby providing direct
evidence for the beyond-Landau nature of metallic quantum criticality. Our
findings also delineate the role of electronic localization
transitions in strange metal phenomena, which appear to be central to the
cuprate superconductors, bilayer graphene, and a variety of other strongly
correlated materials \cite{Kei17.1} and beyond \cite{Bro18.1}.\\

\noindent{\Large\bf Acknowledgments}

\noindent We thank P.\ Gegenwart, Y.-B.\ Kim, H.\ von L\"ohneysen, S.\ Nakatsuji, H.-C.\ N\"agerl, and A.\ Prokofiev for useful discussions.
Financial support for this work was provided by the European Research Council
(ERC Advanced Grant 227378), the U.S.\ Army Research Office (ARO
W911NF-14-1-0496), and the Austrian Science Fund (FWF W1243 and P29279-N27).
X.L.\ and J.K.\ acknowledge financial support from the National Science
Foundation (NSF MRSEC DMR-1720595) and the ARO (W911NF-17-1-0259). Q.S.\
acknowledges financial support from the NSF (DMR-1611392), the Robert A.\ Welch
Foundation (C-1411), and the ARO (W911NF-14-1-0525), and hospitality of the
University of California at Berkeley, the Aspen Center for Physics (NSF grant
PHY-1607611), and the Los Alamos National Laboratory (via a Ulam Scholarship
from the Center for Nonlinear Studies).


\newpage



\begin{figure*}[ht!]

\vspace{-1cm}

{\large\bf\hspace{1.1cm} A \hspace{5cm} B}
\vspace{-0.1cm}

\centering
\subfigure{\hspace{-0.2cm}\includegraphics*[height=0.23\textheight]{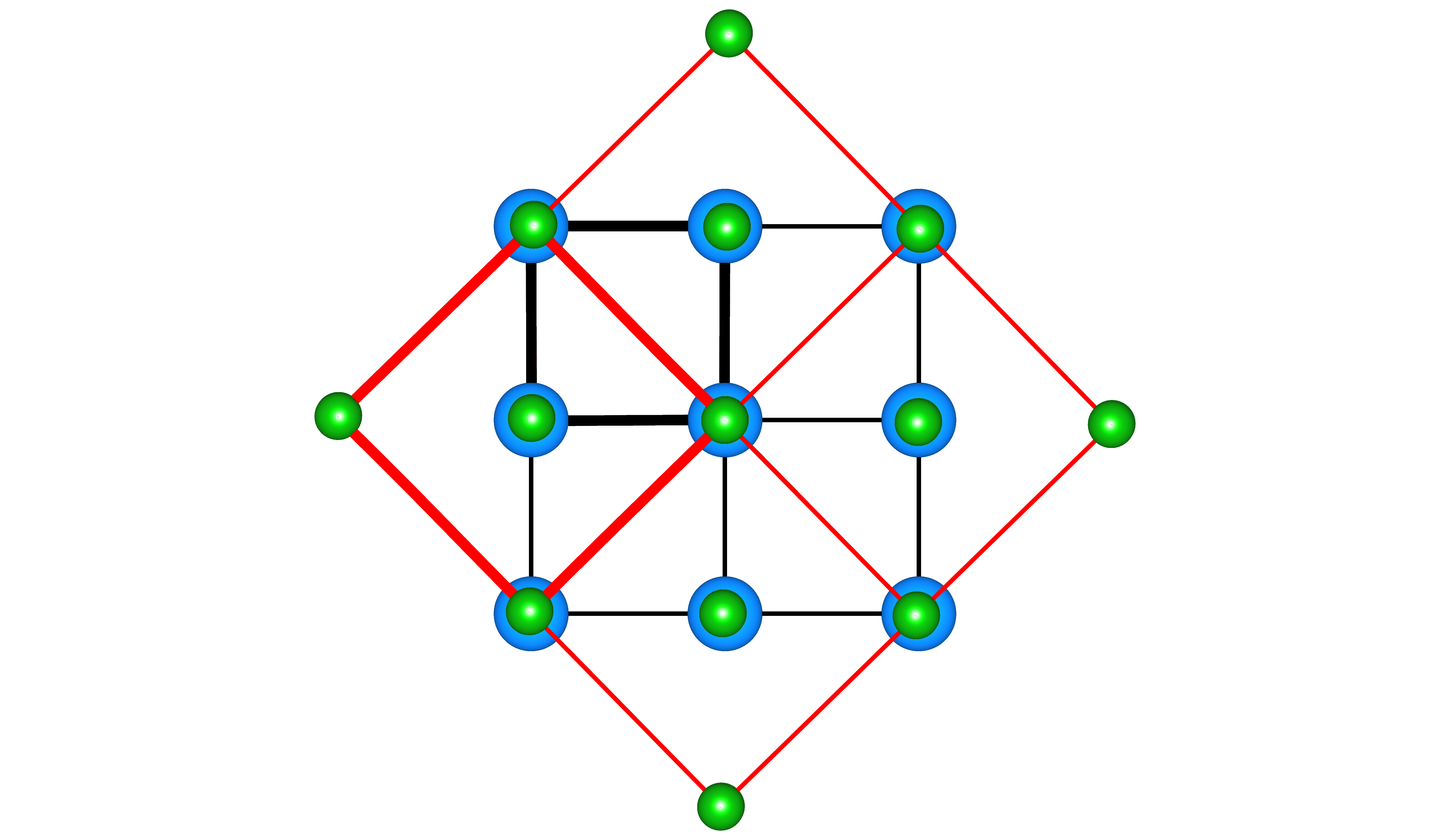}\hspace{-1cm}\includegraphics*[height=0.259\textheight]{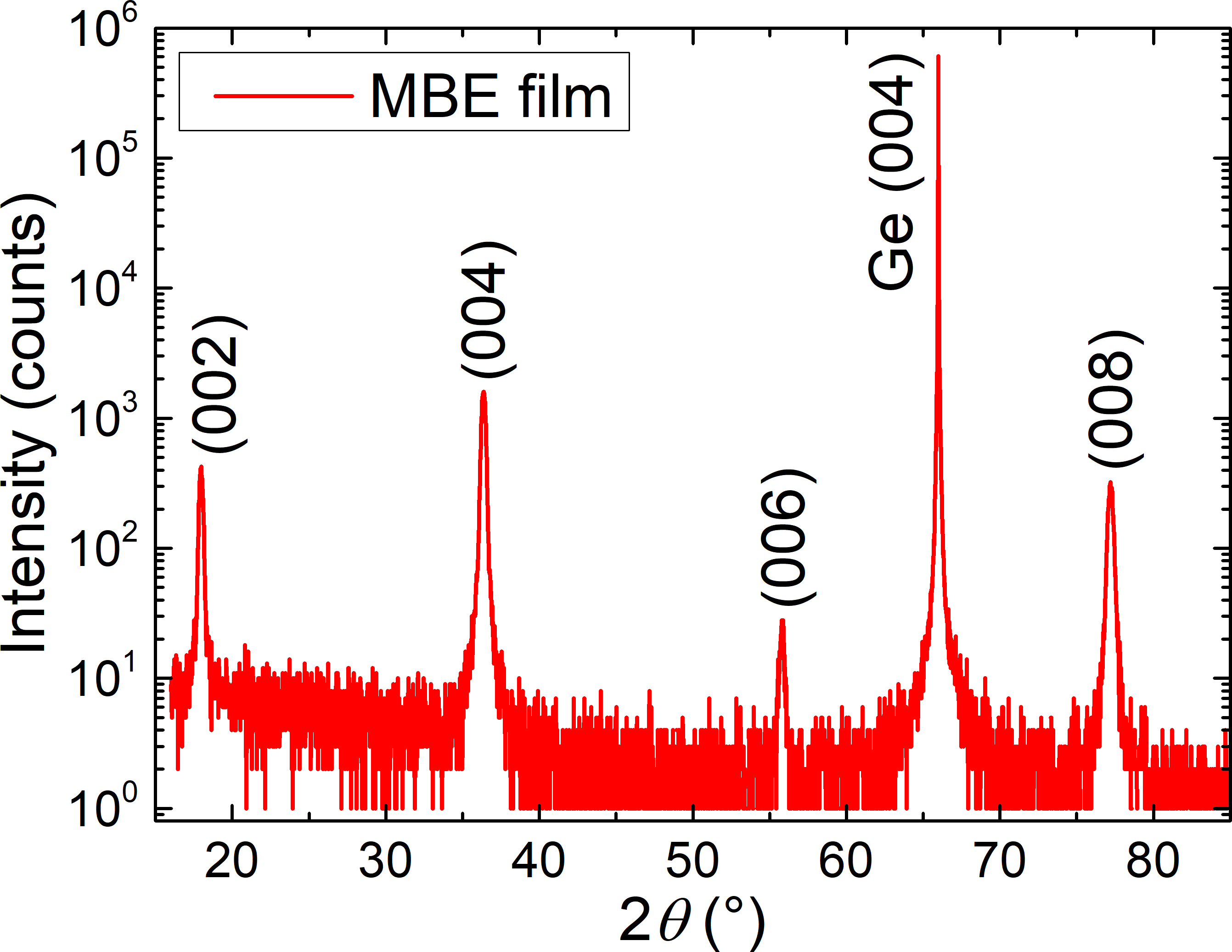}}\hfill
\vspace{0.5cm}

\centering
\subfigure{\hspace{2cm}\includegraphics*[width=0.8\textwidth]{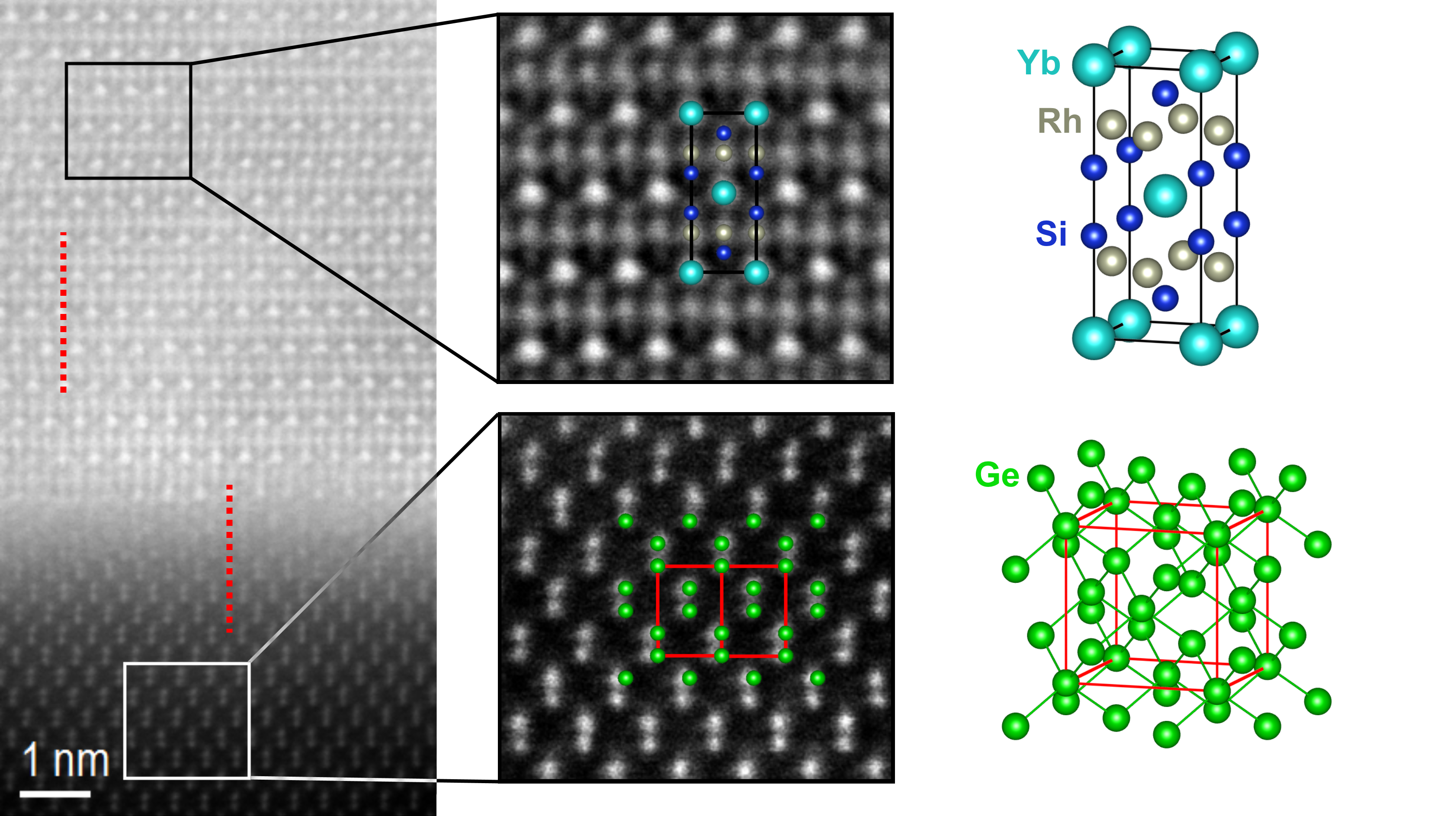}}\hfill
\vspace{0.7cm}

\subfigure{\hspace{-0.0cm}\includegraphics*[height=0.25\textheight]{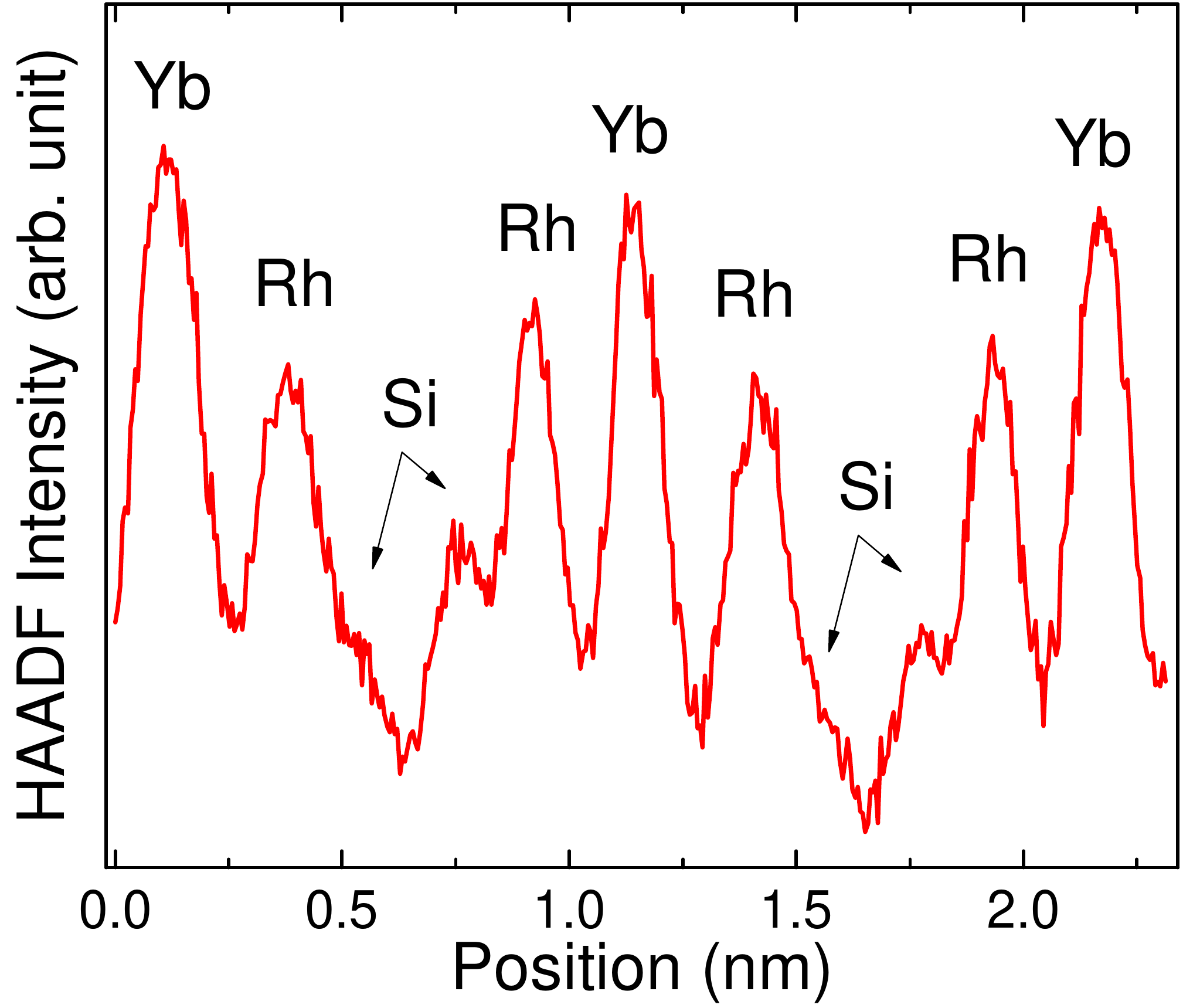}\hspace{0.5cm}\includegraphics*[height=0.25\textheight]{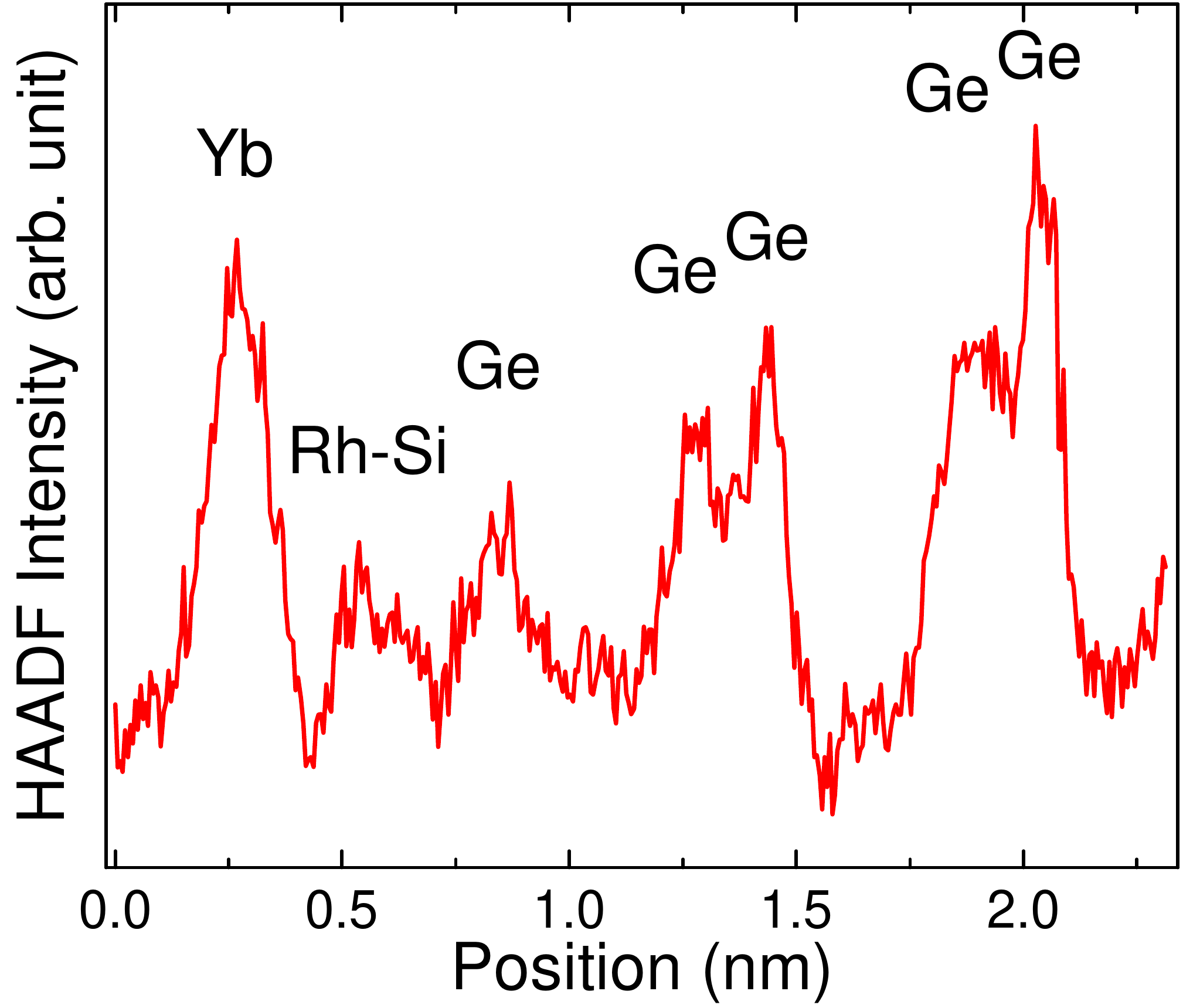}}

\vspace{-14.3cm}

\hspace{-13.1cm}{\large\bf C}
\vspace{7.8cm}

\hspace{-13.1cm}{\large\bf D}

\end{figure*}

\newpage

\begin{figure}[t!]

\caption{\label{MBEfilm}
\baselineskip24pt
{\bf YbRh$_2$Si$_2$ thin films grown by molecular beam epitaxy.}  ({\bf
A}) Visualization of the lattice matching between YbRh$_2$Si$_2$ (blue circles,
black lines) and Ge (green circles, red lines), with the crystallographic $c$
directions pointing out of the plane. For the Yb atoms to associate with the Ge
atoms, the respective unit cells (thick lines, see panel {\bf C}\,right) are
rotated by $45^{\circ}$ with respect to each other around the $c$ direction.
({\bf B}) High-resolution X-ray diffraction pattern, confirming that the film is
phase pure YbRh$_2$Si$_2$. ({\bf C}) Atomic resolution high-angle annular dark
field scanning transmission electron microscopy (HAADF-STEM) image of the
interface between film (top left) and substrate (bottom left), representative
enlarged views with simulated overlays (center), and the corresponding unit
cells in their relaxed states (right). ({\bf D}) Intensity profiles along the
red dashed lines in ({\bf C} left), within the film (upper line, left
panel) and across the interface (lower line, right panel).}

\end{figure}

\clearpage

\newpage


\begin{figure*}[ht!]

\centering
\subfigure{\includegraphics*[width=0.5\textwidth]{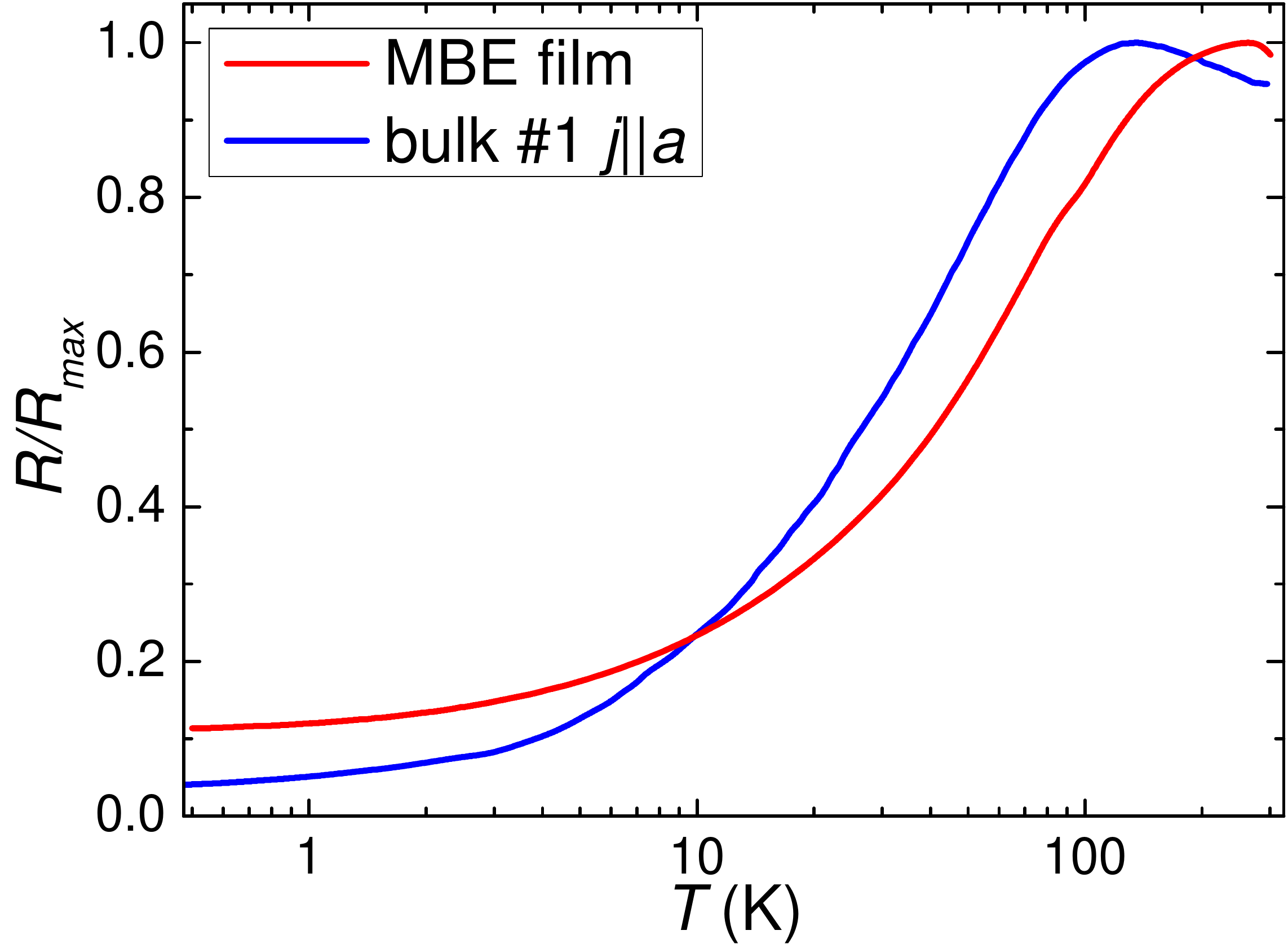}}\hfill

\subfigure{\includegraphics*[width=0.5\textwidth]{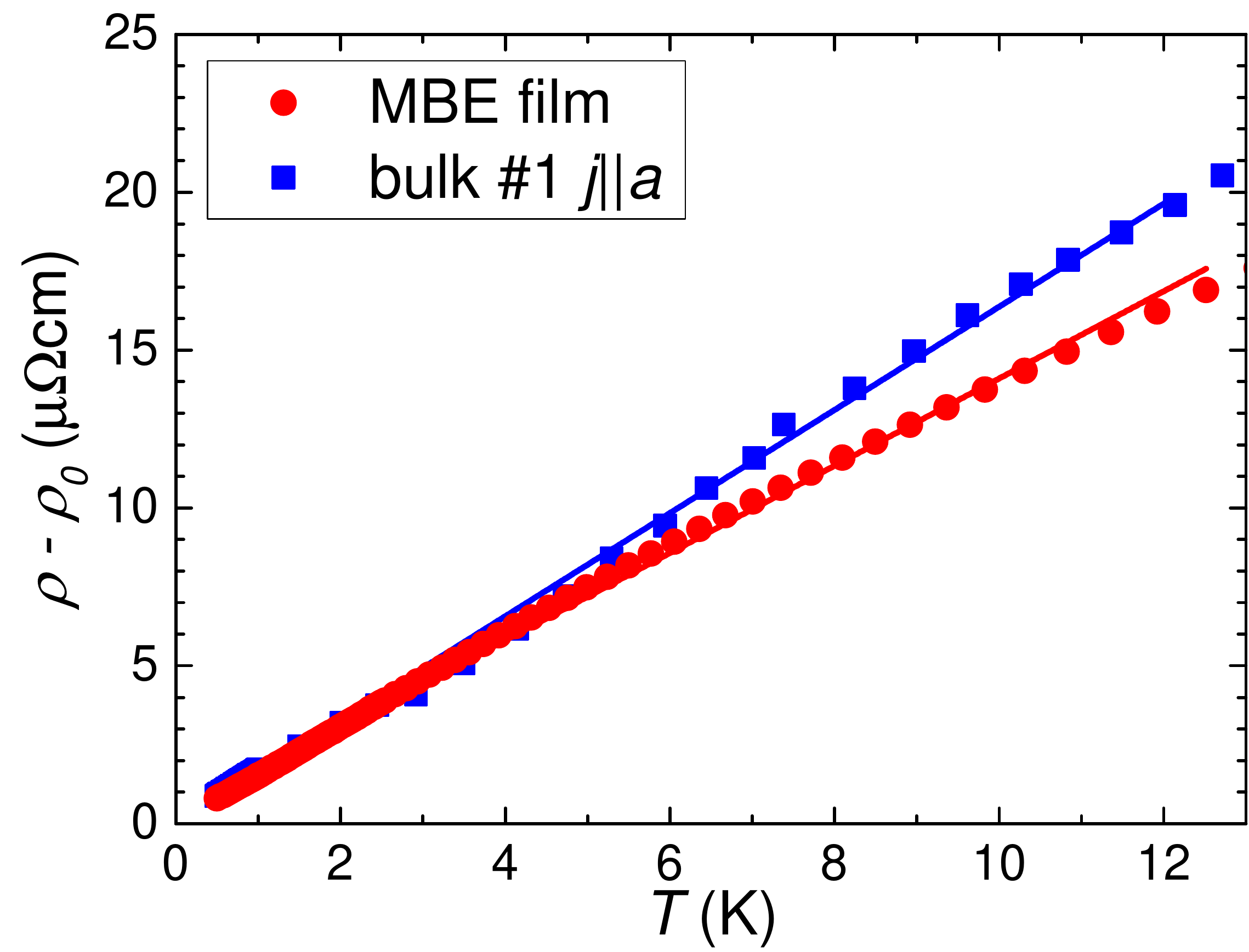}}\hfill

\subfigure{\includegraphics*[width=0.5\textwidth]{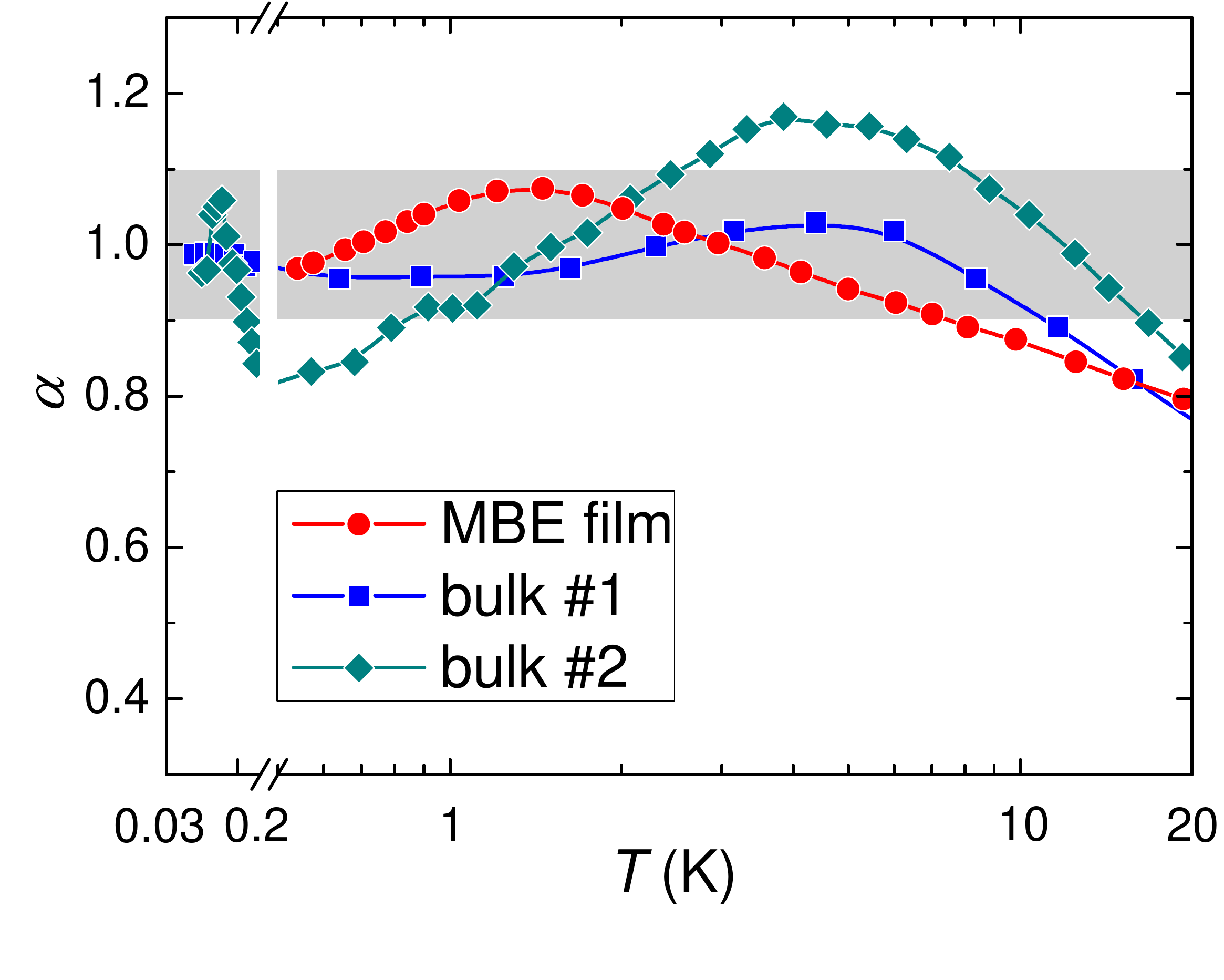}}

\vspace{-20cm}

{\large\bf\hspace{-8cm} A}
\vspace{5.8cm}

{\large\bf\hspace{-8cm} B}
\vspace{5.8cm}

{\large\bf\hspace{-8cm} C}
\vspace{6cm}

\end{figure*}

\newpage

\begin{figure}[t!]

\caption{\label{rhoT}
\baselineskip24pt
{\bf Electrical resistivity of MBE-grown YbRh$_2$Si$_2$.} ({\bf A}) Normalized
resistance of an MBE-grown YbRh$_2$Si$_2$ film, and a bulk single crystal with
current $j$ within the tetragonal $aa$ plane \cite{Tro00.2} for comparison. The
film was measured using the van der Pauw technique. ({\bf B}) Corresponding
low-temperature resistivities, with the residual resistivities ($\rho_0 =
11.6$\,$\mu\Omega$cm and $2.45$\,$\mu\Omega$cm for the MBE and bulk
\cite{Tro00.2} samples, respectively, determined by linear-in-$T$ fits to the
data below 1\,K) subtracted, displaying non-Fermi liquid behavior (lines
represent $\rho - \rho_0 =  A' T^{\alpha}$ fits with constant $\alpha$ to the
data below 12\,K). ({\bf C}) Temperature-dependent exponent $\alpha$, determined
as $\partial \log(\rho-\rho_0)/\partial \log(T)$ for the two above samples, and
an additional bulk single crystal (\#\,2, $\rho_0 = 0.5$\,$\mu\Omega$cm)
\cite{Wes09.1} for comparison. The exponent approaches 1 in the low-temperature
limit, but falls below 0.9 above about 8\,K for our thin film, and above 11 and
16\,K for two different bulk single crystals \cite{Tro00.2,Wes09.1}. Within the
experimental error bar of $\Delta\alpha \approx 10$\%, all data points below
15\,K fall in a zone of $\alpha = 1.0 \pm 0.1$ (grey shading).}

\end{figure}

\clearpage

\newpage


\newpage
    

\end{document}